\documentclass{aa}  
\usepackage{graphicx}
\usepackage{xcolor}
\usepackage{txfonts}

\begin{document}

\title{Numerical solutions of the complete two-body system in QUMOND}
\author{J. Pflamm-Altenburg
          \inst{1}}

   \institute{Helmholtz-Institut f\"ur Strahlen- und Kernphysik,
     Nussallee 14-16,
     D-53115 Bonn 
     \email{jpa@hiskp.uni-bonn.de}
   }

  \abstract
      {Due to the non-linearity of the QUMOND field equations, in the modelling of binaries
        so far the two-body system is replaced by an effective one-body system, where
        the central particle contains the total mass of both binary components and is orbited
        by a massless test particle.}
      {In this work, the discrepancy between the effective one-body treatment and
        the complete two-body solution in QUMOND is quantified.}
      {Particles are treated as limits of Dirac sequences. Then, the QUMOND contribution
        to the total kinematical acceleration of a particle
        is expressed as a Green's integral which is calculated numerically.}
      {In the non-linear transition regime the kinematical acceleration
        of the effective one-body system with a total mass of 2\,$M_\odot$
        is up to a factor
        of 1.44 higher than the Newtonian
        acceleration, whereas the acceleration is only
        boosted by a factor of 1.2--1.3 in the two-body
        system in the case of the simple transition function.}
      {}


   \maketitle

   \section{Introduction}
   
   After the observation that flat rotation curves 
   are a common property of disk galaxies
   \citep[eg.][]{rubin1978a,rubin1980a,bosma1981a}
   it has been realised that this can be easily understood
   if the kinematical acceleration, $a$, and the gravitational acceleration, $g$,
   are identical in regions where the accelerations are above a common
   threshold, $a_0$, and the kinematical acceleration is proportional to
   the square root of the
   gravitational acceleration in low acceleration
   regimes 
   \citep{milgrom1983b,milgrom1983c,milgrom1983a},
   \begin{equation}\label{eq_a_g_cond}
     a = g\;\mathrm{for}\;a\gg a_0\;\mathrm{and}\;
     a \propto \sqrt{g}\;\mathrm{for}\;a\ll a_0\,.
   \end{equation}
   This modification of Newtonian dynamics (MOND) immediately links 
   the total baryonic mass of a galaxy, $M_\mathbf{b}$, to the  asymptotic
   flat rotation curve with velocity, $v_\infty$, by $v_\infty^4 = a_0 G M_\mathbf{b}$
   \citep{milgrom1983b} which has been confirmed later to high degree
   observationally \citep{mcgaugh2000a}.
   
   MOND field theories have been formulated as an extension of the Poisson
   equation. The AQUAL-formulation keeps the source term of the Poisson
   equation and replaces the Laplace operator by a non-linear operator
   in order to fulfil the boundary behaviour of the acceleration field
   in the Newtonian and MONDian regime \citep{bekenstein1984a}.
   In a second widely used formulation (QUMOND) the Laplace operator of the
   Poisson equation is kept and the source term is replaced by a
   non-linear expression \citep{milgrom2010a}.

   The transition from the Newtonian to the MONDian regime
   is generally described by scaling the acceleration with a
   mapping where the argument is the ratio of the local kinematical or gravitational
   acceleration and $a_0$.
   However, there is no logical constraint for this transition function
   to be a function of $a/a_0$ only.
   The form of the transition
   can also for example depend on the local mass density and/or on the
   local symmetry of the mass distribution, meaning a dependence on $\rho$
   and $\nabla\rho$. The only boundary conditions that MOND imposes are those in
   Eq.~(\ref{eq_a_g_cond}) as already pointed out by \citet{bekenstein1984a}.
   
   If MOND-type theories are a valid description of gravity in the weak field
   limit then small deviations from Newtonian gravity are expected
   already on sub-Galactic scales
   \citep{milgrom1983b,milgrom1983c,milgrom1983a}.

   Open star clusters in the solar vicinity show asymmetric tidal tails
   \citep{jerabkova2021a,boffin2022a}, which are unlikely to form in
   Newtonian dynamics \citep{pflamm-altenburg2023a,kroupa2024a}
   but appear in QUMOND dynamics
   \citep{kroupa2022a} and Milgrom-law-dynamics (MLD)
   \citep{pflamm-altenburg2025a}.

   MONDian dynamics is expected to show observable
   differences to Newtonian dynamics
   in the orbital evolution of  extreme trans-Neptunian objects
   \citep{pauco2016a,pauco2017a,brown2023a}.

   \citet{blanchet2011a} quantified the additional quadrupole
   moment produced by MOND in the inner solar system and concluded
   that the MOND interpolation function for the Solar system
   requires a very sharp transition
   in order be in agreement with orbital precession data of Jupiter.
   The requirement of a rapid transition from the Newtonian to the MONDian
   regime is also inline with the results from Cassini tracking data
   \citep{hees2014a,hees2016a}.
   
   \citet{hernandez2012a} proposed to analyse the kinematics
   of wide binaries as a critical test. Local binaries with a
   separation of more than ${\approx}\,7000\,\rm AU$ should have
   internal accelerations and relative velocities slightly higher
   than expected in Newtonian dynamics.

   Some subsequent studies have reported a non-Newtonian dynamical
   behaviour of wide binaries \citep{hernandez2019a,hernandez2022a,
     hernandez2023a,hernandez2024a,chae2023a,chae2024a,chae2024b},
   while others found no deviation from Newtonian expectation
   \citep{pittoris2023a,banik2024a,cookson2024a}.
   However, \citet{hernandez2024b} and \citet{hernandez2025a}
   reported the discovery of critical problems
   in the statistical analysis of wide binaries in
   \citet{pittoris2023a}, \cite{banik2024a} and \citet{cookson2024a}.

   In addition to the critics mentioned above the 
   \citet{banik2024a} study suffers from an additional problem.
   The orbital modelling of wide binaries in \citet{banik2024a}
   is done by replacing the two-body system by an effective
   one-body system, where the central mass is equal to the sum of
   both binary components
   and is orbited by a massless test
   particle \citep[Sect. 3.1.1., 4th par.]{banik2024a}.
   They argue that in the inner region the binaries follow
   Newtonian dynamics and in the outer regions
   the binaries follow an effective Newtonian dynamics where
   the scaling factor is dominated by the larger
   external acceleration (called the external field effect, EFE).
   All wide binaries are treated to lie either in
   the Newtonian or the EFE-dominated regime and apply linearity
   between the mass distribution and the resulting gravitational
   fields.

   However, the FWHM of the mass distribution of the stellar
   masses of the wide binary sample lies in the range from 0.4 to 1~$M_\odot$
   \citep[Fig.~5]{banik2024a} and the vast majority of wide binaries
   have projected separations of 5000~AU and less
   \citep[Fig.~6]{banik2024a}. At 2000~AU distance
   from a 1~$M_\odot$ the internal Newtonian acceleration is $12.6\,a_0$ and at
   5000~AU distance
   $2\,a_0$. For a $0.4\,M_\odot$ star the acceleration
   is $5\,a_0$ at 2000~AU and $0.8\,a_0$ at 5000~AU. As their assumed external
   acceleration exerted by the Galactic gravitational field
   is about $1.14\,a_0$, most of the binaries are located
   in the non-linear transition regime from Newtonian to EFE-MOND dynamics. 
   \citet{banik2024a} claim that the orbital evolution
   of wide binaries underlies
   MOND is excluded by a 16~$\sigma$ confidence. However, \citet[][Sect.~3.2]{hernandez2024b}
   (among other inconsistencies) points out that the sample used in \citet{banik2024a}
   contains kinematic contaminants and that in the case of the binary subsample with
   a separation between 12 and 5 kAU their MOND models are closer to the observations
   than the Newton models. 
   Furthermore, claiming the failure
   of MOND with a such high confidence requires
   a proper and very accurate orbit modelling in MOND on the theoretical side.
   Therefore, the strong conclusion made by
   \citet{banik2024a} is very questionable.

   The main motivation for this work is therefore to calculate
   complete solutions of the
   two-body problem in MONDian field theories and the comparison with the
   effective one-body problem. As a first step the QUMOND formulation is
   considered in this work due to the linearity of the Laplace operator.
   In Sect.~\ref{sec_qumond_nbody} an integral expression for the
   acceleration of a point mass in an
   $N$-body system is calculated using the Green's integral representation
   of the solution of the linear QUMOND pde. Section~\ref{sec_numerics}
   summarises
   the numerical routines used for the calculation of the integrals. 
   In Sect.~\ref{sec_qumond_isolation}
   the Green's integrals are calculated numerically for an isolated binary
   and compared with analytical expressions for the deep MOND case.
   The case of a collinear
   binary with a total mass of 2\,$M_\odot$
   embedded in an external Galactic field is considered
   in Sect.~\ref{sec_qumond_efe}. In Sect.~\ref{sec_3d} this binary is
   analysed in three dimensions.
   Section~\ref{sec_mld} explores an embedded binary in Milgrom law dynamics
   (MLD)
   in brief followed by a proposal of an alternative observational test in
   Sect.~\ref{sec_proposal}.

   \section{The particle acceleration of an $N$-body system in QUMOND}
   \label{sec_qumond_nbody}
   The QUMOND formulation was introduced by \citet{milgrom2010a}.
   The kinematical
   potential, $\Phi_\mathrm{Q}$, determines the acceleration field of a fluid
   by
   \begin{equation}\label{eq_delta_phi_Q}
     \mathbf{a} = -\nabla\Phi_\mathrm{Q}
   \end{equation}
   and is linked to a non-linear source,
   \begin{equation}\label{eq_qumond}
     \Delta\Phi_\mathrm{Q} = \nabla(\nu(|\Phi_\mathrm{N}|/a_0)\,
     \nabla\Phi_\mathrm{N})\,,
   \end{equation}
   where $\Phi_\mathrm{N}$ is the Newtonian potential and obtained
   from the standard Poisson
   equation,
   \begin{equation}\label{eq_poisson}
     \Delta\Phi_\mathrm{N} = 4\pi G \rho\,,
   \end{equation}
   with the baryonic mass density, $\rho$.

   The $\nu$-function joins the MONDian and the Newtonian regime continuously
   and is currently based on observations of rotation curves of disk galaxies. 
   The choice of the transition
   function is restricted by the boundary conditions in
   Eq.~(\ref{eq_a_g_cond}) requiring
   \begin{equation}\label{eq_nu_cond}
     \nu(y) = 1\;\mathrm{for}\;y\gg 1\;\mathrm{and}\;
     \nu(y) = y^{-1/2}\;\mathrm{for}\;y\ll 1\;.
   \end{equation}
   The transition functions used in this work are summarised in
   Table~\ref{tab_nu_fcts}. The threshold acceleration is set
   to $a_0=1.2\times 10^{-10}\,\rm m/s^2$ \citep{begeman1991a} corresponding to
   $a_0 = 7.988\times 10^{-7}\,\rm AU/yr^{-2}$.

   \begin{table}
     \caption{\label{tab_nu_fcts}QUMOND transition functions used in this work.}
     \begin{tabular}{cc}
       name & function\\
       \hline
       simple   & $\nu_\mathrm{sim}(y) = \frac{1+\sqrt{1+4/y}}{2}$ \\
       standard & $\nu_{std}(y) = \sqrt{\frac{1+\sqrt{1+4/y^2}}{2}}$  \\
       McG08 & $\nu_\mathrm{McG08}(y) = \left(1-e^{-\sqrt{y}}\right)^{-1}$ \\
       maximum & $\nu_\mathrm{max}(y) = \left\{\begin{array}{c@{,}c}
       1 \;\;&\;\;y \ge1\\
       y^{-1/2} \;\;&\;\;y <1\\
       \end{array}\right.$\\
     \end{tabular}
    
     \tablefoot{The simple transition function
       was introduced by \citet{famaey2005a} in its corresponding
       AQUAL form and used in the wide-binary
       analysis in \citet{banik2024a}. The standard interpolation function
       was introduced by \citet{kent1987a} in its corresponding
       AQUAL form. The McG08 transition function was introduced by
       \citet{mcgaugh2008a}. The maximum transition
       function is introduced here 
       for debugging purposes and as a convenient indicator whether a region
       is Newtonian or MONDian.
       }
   \end{table}
   
   As already mentioned in \citet{milgrom2010a} the solution of the
   kinematical potential is given by the Green's integral expression
   with respect to an arbitrary but constant reference point, $\mathbf{x}_0$, 
   \begin{multline}\label{eq_pot_green_qumond}
     \Phi_\mathrm{Q}(\mathbf{x}) =\\
     \int_{\mathbb{R}^3}
     \left(G(\mathbf{x},\mathbf{y})-G(\mathbf{x}_0,\mathbf{y})\right)
  \;\nabla\left[\nu(|\Phi_\mathrm{N}(\mathbf{y})|/a_0)\,\nabla\Phi_\mathrm{N}(\mathbf{y})\right]
  d^3\mathbf{y}\,,
   \end{multline}
    where
   \begin{equation}
     G(\mathbf{x},\mathbf{y}) = -\frac{1}{4\pi|\mathbf{x}-\mathbf{y}|}
   \end{equation}
   is the Green's function of the Laplace operator in the case of an
   unbounded domain.

   A practical equivalent formulation to Eq.~(\ref{eq_qumond}) is
   \begin{equation}
     \Delta\Phi_\mathrm{Q}=4\pi G\left(\rho+\rho_\mathrm{pdm}\right)\,,
   \end{equation}
   where $\rho_\mathrm{pdm}$ is the so-called phantom dark matter
   \begin{equation}
     \rho_\mathrm{pdm}:=\frac{1}{4\pi G}
     \nabla\left[\left(\nu(|\nabla\Phi_\mathrm{N}|/a_0)-1\right)\nabla\Phi_\mathrm{N}\right]\;.
     \end{equation}
   The phantom dark matter can be interpreted as an additional but non existing
   matter distribution causing the dynamical deviation of a system from an
   evolution in pure Newtonian dynamics.

   The acceleration field at an arbitrary position, $\mathbf{x}$,
   can be calculated from the gradient of the kinematical potential
   with 
\begin{equation}
  \mathbf{a}(\mathbf{x})
  = -\int_{\mathbb{R}^3}\mathbf{G}(\mathbf{x},\mathbf{y})
  \;\nabla\left[\nu(|\Phi_\mathrm{N}(\mathbf{y})|/a_0)\,\nabla\Phi_\mathrm{N}(\mathbf{y})\right]
  d^3\mathbf{y}
  \;,
\end{equation}
where $\mathbf{G}(\mathbf{x},\mathbf{y})$ is the gradient
(with respect to $\mathbf{x}$)
of the Green's function
\begin{equation}
  \mathbf{G}(\mathbf{x},\mathbf{y})=
  \frac{1}{4\pi}\frac{\mathbf{x} - \mathbf{y}}{|\mathbf{x}-\mathbf{y}|^3}\,.
\end{equation}

Applying the product rule to the divergence operator the integral for
the acceleration can be split into two terms
\begin{multline}\label{eq_a_qumond}
  \mathbf{a}(\mathbf{x}) =-
  \int_{\mathbb{R}^3}\mathbf{G}(\mathbf{x},\mathbf{y})
  \nabla\left[\nu(|\Phi_\mathrm{N}(\mathbf{y})|/a_0)\right]\bullet
  \,\nabla\Phi_\mathrm{N}(\mathbf{y})\;
  d^3\mathbf{y}\\
  \phantom{\mathbf{a}(\mathbf{x})}
  -4\pi G\int_{\mathbb{R}^3}\mathbf{G}(\mathbf{x},\mathbf{y})\nu(|\Phi_\mathrm{N}(\mathbf{y})|/a_0)\;\rho(\mathbf{y})\;
  d^3\mathbf{y}\,.\hfill
\end{multline}
So far, the mass density distribution can be arbitrary.

Now, consider a system of $n$ particles with masses ($m_1,\ldots,m_n$)
and position vectors $(\mathbf{r}_1,\ldots,\mathbf{r}_n)$. The corresponding
mass density distribution is
\begin{equation}\label{eq_mass_density}
  \rho(\mathbf{y}) =\sum_{i=1}^n m_i\delta(\mathbf{y}-\mathbf{r}_i)\,.
\end{equation}
The Newtonian acceleration field at a position $\mathbf{x}\not=\mathbf{r}_i$
between the particles is given by
\begin{equation}
  \mathbf{g}(\mathbf{x})
  =-4\pi G\int_{\mathbb{R}^3}\mathbf{G}(\mathbf{x},\mathbf{y})\;\rho(\mathbf{y})\;
  d^3\mathbf{y}=-G \sum\limits_{i=1}^n m_i \frac{\mathbf{x}-\mathbf{r}_i}{|\mathbf{x}-\mathbf{r}_i|^3}\,.
\end{equation}
The Newtonian acceleration of the particle $j$ can be obtained by evaluating the
integral for $\mathbf{x}=\mathbf{r}_j$,
\begin{multline}
  \mathbf{g}_j
  =-4\pi G\int_{\mathbb{R}^3}\mathbf{G}(\mathbf{r}_j,\mathbf{y})\;\rho(\mathbf{y})\;
  d^3\mathbf{y}\\
  =-G \sum\limits_{i=1}^n m_i
  \int_{\mathbb{R}^3}\mathbf{G}(\mathbf{r}_j,\mathbf{y})
  \delta(\mathbf{y}-\mathbf{r}_i)\;d^3\mathbf{y}\,.
  \hfill
\end{multline}
Treating a point mass as the limit of a symmetric Dirac  sequence
$(\delta_\varepsilon)_{\varepsilon>0}$ with compact support identical to
the ball with radius $\varepsilon$ and centre $\mathbf{r}_j$
leads to
\begin{multline}
  \mathbf{g}_j
  =-G  m_j
  \lim\limits_{\varepsilon\to 0}\int_{\mathbb{R}^3}\mathbf{G}(\mathbf{r}_j,\mathbf{y})
  \delta_\varepsilon(\mathbf{y}-\mathbf{r}_j)\;d^3\mathbf{y}
\\
  -G \sum\limits_{\substack{i=1\\i\not=j}}^n m_i
  \lim\limits_{\varepsilon\to 0}\int_{\mathbb{R}^3}\mathbf{G}(\mathbf{r}_j,\mathbf{y})
  \delta_\varepsilon(\mathbf{y}-\mathbf{r}_i)\;d^3\mathbf{y}\,.\hfill
\end{multline}
The integrand of the first term is antisymmetric with a symmetric support
and vanishes for all $\varepsilon$. Thus, the self contribution to the
acceleration is 0. The integrand of the second term does not contain
a singularity and therefore converges to the value of the Green's function
at the respective particle position.
The Newtonian gradient of the whole (discontinuous) acceleration field
is then given by the general expression
\begin{equation}
  \nabla\Phi_\mathrm{N}(\mathbf{x})
  =
  \left\{
\begin{array}{ccc}
  \sum\limits_{i=1}^{n} Gm_i\frac{\mathbf{x}-\mathbf{r}_i}{|\mathbf{x}-\mathbf{r}_i|^3}
  & , &\mathbf{x}\not= \mathbf{r}_j\\
  &&\\
  \sum\limits_{\substack{i=1\\i\not= j}}^{n} Gm_i\frac{\mathbf{r}_j-\mathbf{r}_i}{|\mathbf{r}_j-\mathbf{r}_i|^3}
  & , & \mathbf{x} = \mathbf{r}_j\\
  \end{array}
  \right.\, .
\end{equation}

In order to calculate the acceleration of the $j$th particle
in QUMOND, we set
$\mathbf{x} = \mathbf{r}_j$ in Eq.~(\ref{eq_a_qumond}).
The delta-distribution in the mass density of a particle system
in Eq.~(\ref{eq_mass_density})
concentrates the contribution to the integral of the second term
in Eq.~(\ref{eq_a_qumond}) to the vicinity of
the singularities, where the transition function tends to unity.
Thus, the second term in Eq.~(\ref{eq_a_qumond}) is identical to the
Newtonian particle acceleration.

The QUMOND particle acceleration can then be expressed by
\begin{equation}\label{eq_g_alpha}
  \mathbf{a}_j = \boldsymbol{\alpha}_j + \mathbf{g}_j\,,
\end{equation}
where $\boldsymbol{\alpha}_j$ is the first term in Eq.~(\ref{eq_a_qumond}).

Applying the chain rule to the $\nu$-term in the integral for
$\boldsymbol{\alpha}_j$ we obtain
\begin{equation}\label{eq_alpha}
  \boldsymbol{\alpha}_j = -
  \int\limits_{\mathbb{R}^3}
  \mathbf{G}(\mathbf{r}_j,\mathbf{y})
  \frac{\nu^\prime(|\nabla\Phi_\mathrm{N}(\mathbf{y})|/a_0)}
  {|\nabla\Phi_\mathrm{N}(\mathbf{y})|a_0}
  \left(\nabla\Phi_\mathrm{N}\hat{\mathbf{T}}\right)(\mathbf{y})
  \bullet\nabla\Phi_\mathrm{N}(\mathbf{y})
  \;d^3\mathbf{y}\,,
\end{equation}
where 
\begin{equation}
  \left(\nabla\Phi_\mathrm{N}\hat{\mathbf{T}}\right)_k(\mathbf{y}):=
  \sum\limits_{l=1}^{3}\left(\nabla\Phi_\mathrm{N}\right)_l(\mathbf{y})
  \hat{\mathbf{T}}_{l,k}(\mathbf{y})\;,\;1\le k\le 3
\end{equation}
and
\begin{equation}
  \hat{\mathbf{T}}_{l,k}(\mathbf{y}) = \partial_{l}\partial_{k}\Phi_\mathrm{N}(\mathbf{y})=
  \sum\limits_{i=1}^{n}
  \frac{Gm_i}{|\mathbf{y}-\mathbf{r}_i|^3}
  \left(\delta_{kl}-3\frac{(y_k-r_{i,k})(y_l-r_{i,l})}{|\mathbf{y}-\mathbf{r}_i|^2}\right)\,
\end{equation}
is the symmetric $3\times 3$ tidal tensor in the case of a Newtonian
point mass system.

If the transition function tends fast enough to unity
when approaching the particles, the integrand tends to zero and the
particles are surrounded by phantom dark matter free bubbles.
Thus, the integral of the the QUMOND contribution is free of any singularity
and can be calculated conveniently.

The phantom dark matter density of an $N$-body system is given by
\begin{equation}
  \rho_{\mathrm{pdm}}(\mathbf{y}) = \frac{1}{4\pi G}
  \frac{\nu^\prime(|\nabla\Phi_\mathrm{N}(\mathbf{y})|/a_0)}{|\nabla\Phi_\mathrm{N}(\mathbf{y})|a_0}
  \left(\nabla\Phi_\mathrm{N}\hat{\mathbf{T}}\right)(\mathbf{y})
  \bullet\nabla\Phi_\mathrm{N}(\mathbf{y})\,.
\end{equation}
Thus, at large distances from a point mass or from the complete $N$-body system
the phantom dark matter density behaves as
\begin{equation}\label{eq_pdm_distant}
  \rho_{\mathrm{pdm}}(\mathbf{y}) \approx \frac{1}{4\pi}\,
  \sqrt{\frac{a_0 m}{G}}\frac{1}{r^2}\,,
\end{equation}
where $r$ is the distance to the point mass or the mass concentration.

One important point in Eq.~(\ref{eq_a_qumond}) is
that the second contribution is identical to the Newtonian acceleration
only in the case of a point mass system.
Consider the case where
the point mass system is replaced by a smooth mass density distribution
with identical total mass which is internally MONDian. In the case of a MONDian
smooth mass distribution the $\nu$-function in the second integral has a
larger than unity contribution to the integrand.
The integrand of the first integral contains Newtonian holes around
each particle, if $\nu$ tends to one and its gradient tends to zero
faster than the Newtonian gradient tends to infinity
(depending on the choice of $\nu$).
It is not clear how a change in the mass distribution varies the
values of the two integrals and whether or not
the changes compensate each other.
Here, the external field of the Galaxy is represented by a central 
massive third particle and not by a smooth  external  field.

\section{Integrability of point mass systems}
\begin{figure}
      \includegraphics[width=\columnwidth]{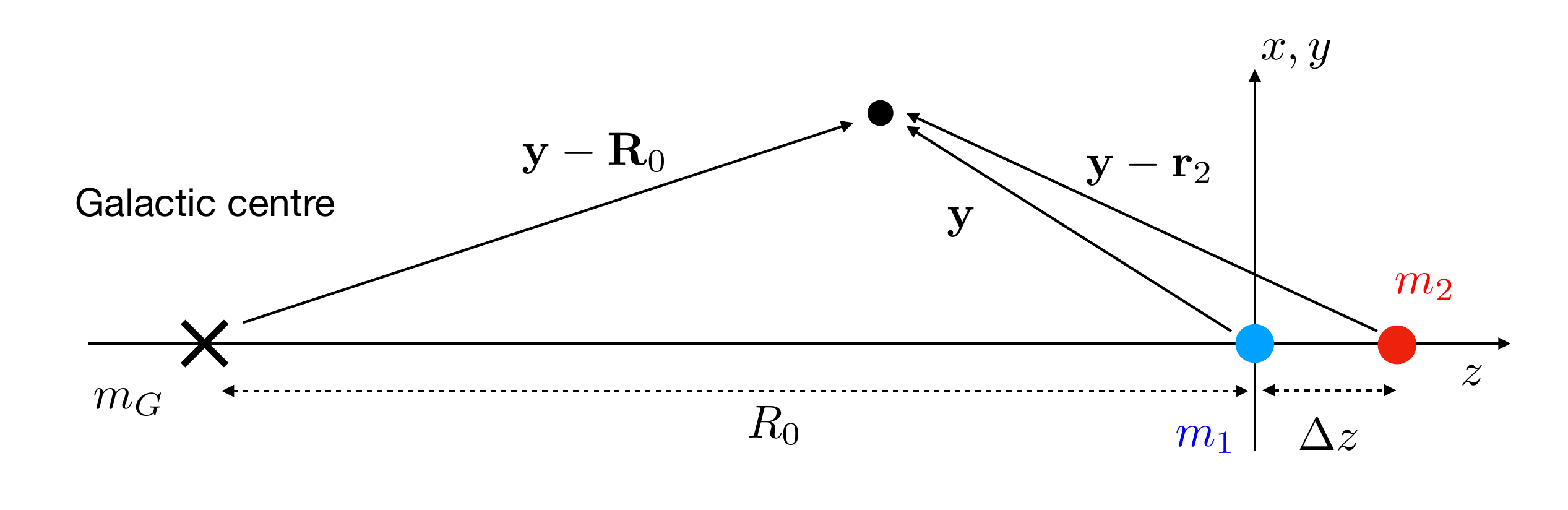}
      \caption{\label{fig_setup}Set up of particle positions.
      }
    \end{figure}

Even if the transition functions are continuous there are three
critical points where the integrability has to be checked.
Figure~\ref{fig_setup} shows the general set up of the
collinear particle system.
The system is shifted such that
    the particle which QUMOND-acceleration
    (Eq.~(\ref{eq_alpha})) has to be calculated is at the origin
    if not stated otherwise.
    The other component of the binary is put on the positive
    $z$-axis with a separation of $\Delta z$ to the first particle.
    In the case of an isolated binary the point mass representing the
    Galaxy is omitted.

\begin{figure}
      \includegraphics[width=\columnwidth]{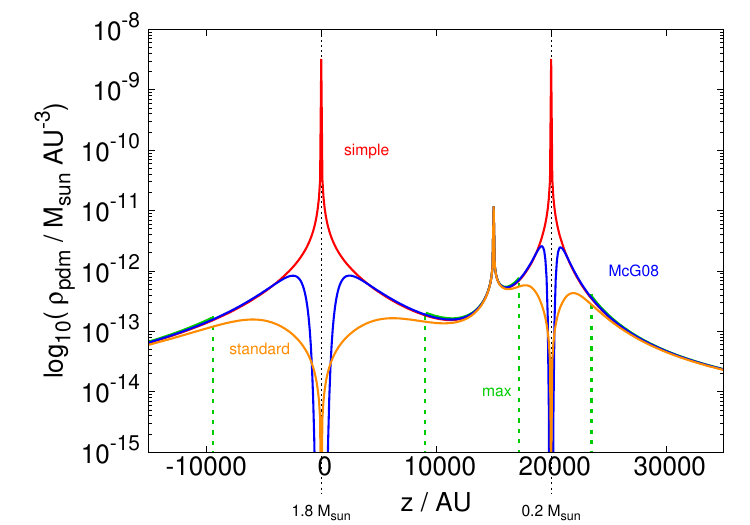}
      \caption{\label{fig_pdm_iso}
        Phantom dark matter density distribution of an isolated two-body system.
      }
    \end{figure}

    The distribution of the phantom dark matter along 
    the connecting line ($z$-axis) of two particles with masses
    $m_1=1.8\,M_\odot$ and $m_2=0.2\,M_\odot$ and a separation of
    $\Delta z = 20\,\rm kAU$ can be seen in Fig.~\ref{fig_pdm_iso}
    for four different transition functions (Table~\ref{tab_nu_fcts}).
    The maximum $\nu$-function indicates the transition from the Newtonian
    to the MONDian regions ($|g_z| = a_0$) at
    $z=-9484\,\rm AU,\;9066\,AU,\;17243\,AU,\;21000\,AU$.
    Outside the Newtonian region the phantom dark matter density distributions
    of all four transition functions converge
    and scale with $r^{-2}$ (Eq.~(\ref{eq_pdm_distant})).

    Between the two particles a peak in the phantom dark matter density
    appears at $z=15\,\rm kAU$ indicating the location of zero-g, where
    the total Newtonian acceleration vanishes, independent of the
    choice of the transition function. It might be worth to explore
    whether or not a real particle can be trapped by this imaginary
    concentration of phantom dark matter.
    The appearance of MOND effects at the location of vanishing gravitational
    forces has already been mentioned by \citet{hernandez2017a}
    in the context of mocking a central black hole in Globular clusters.

    The jump of the phantom dark matter density in the case of  the
    maximum transition function marks the positions where the local
    acceleration is identical to $a_0$. As the tidal tensor can be different
    at these positions the phantom dark matter density can be different
    here, too.

    \subsection{Integrability for $R\to \infty$}
    In order to have integrability for $R\to\infty$, the composition
    $\nu^\prime\circ|\nabla\Phi_\mathrm{N}|)(R)$ needs to scale with $R^{\alpha}$
    with $\alpha \le 3$. As all transition functions, $\nu(y)$,
    scale with $y^{-1/2}$ in the deep MOND regime,
    the composition scales with $R^3$ and integrability
    is ensured for all transitions functions.

    \subsection{Integrability at zero-g points}
    Consider a small ball with radius $\varepsilon$ centred on a zero-g
    point. The Green's function has a nearly constant value.
    The Newtonian acceleration scales
    with $\nabla\Phi_\mathrm{N}\propto\varepsilon$ in the direction of the $z$-axis.
    The whole integrand
    scales by $\varepsilon^{-1/2}$ which is compensated by $\varepsilon^2$
    from the volume element. Thus, integrability
    is ensured for all transitions functions.

    \subsection{Integrability at particle locations}
    Consider a small ball with radius $\varepsilon$ centred on a
    particle. The Green's function scales with $\varepsilon^{-2}$
    if the central particle is the particle whose acceleration
    has to be calculated and is constant in the case of any other particle.
    Thus, integrability needs to be checked only at the particle of interest.
    In this case the composition
    $\nu^\prime\circ|\nabla\Phi_\mathrm{N}|)(\varepsilon)$ needs to scale with
    $\varepsilon^\alpha$ with $\alpha \ge 5$. In the case of the standard
    integration function the composition scales with $\varepsilon^{6}$ and
    integrability is guaranteed.

    Contrary, in the case of the simple transition function the composition
    scales only with $\varepsilon^{4}$. Thus, the integral does not converge.
    This means, the simple transition function can not be used
    in point mass descriptions in QUMOND. However, with respect to the
    particle (or put to the origin) the integrand in Eq.~(\ref{eq_alpha})
    behaves antisymmetric in the vicinity of the particle
    ($\propto \varepsilon^{-3}$). As the domain is symmetric 
    the antisymmetric parts cancel each other if the integral is evaluated
    in the sense of the Cauchy principal value. The remaining symmetric
    part of the integrand is of order $\propto \varepsilon^{-2}$ and is
    integrable.

    In the case of the McGaugh-transition function the composition
    $\nu^\prime\circ|\nabla\Phi_\mathrm{N}|)(\varepsilon)$ can be continued
    at $\varepsilon = 0$ by zero to be $C^{\infty}$. As all derivatives are
    zero it is not analytic and can not be Taylor expanded
    at $\varepsilon=0$.  Switching to spherical coordinates in
    the vicinity of the particle leads to an one-dimensional integral
    of a function proportional to
    $e^{1/\varepsilon} / \left(\varepsilon^5\,\left(e^{1/\varepsilon}-1\right)^2\right)$
    which tends to 0 for $\varepsilon\to 0$.
    Thus integrability is ensured.

\section{Numerical integration}\label{sec_numerics}
For simplicity, we here focus in the first step on the
embedded case where both binary constituents and the
Galaxy point mass are located on one line. Due to rotational symmetry
around the $z$-axis the dimension
of the integral can be reduced to two after switching to polar coordinates
and the calculations are reduced to the integration of the $z$-component
because $a_x=a_y=0$.

    In the case of an isolated binary the integral with
    infinite domain is transformed to a finite domain by the
    mapping
    \begin{multline}
      \psi:\, [0,1[\;\times\;]-1,1[\;\to\;
          \mathbb{R}_{\ge 0}\times\;\mathbb{R}\\
          \hspace{1.8cm}(\xi_r,\xi_z) \mapsto
          \left(\frac{\lambda\xi_r}{1-\xi_r},\frac{\lambda\xi_z}{1-\xi_z^2}\right)\,,
          \hfill
    \end{multline}
    where $\lambda$ is a scaling parameter and set to $\lambda=2 \Delta z$
    in order to resolve the interior of the binary.
    The corresponding Jacobi determinant is
    \begin{equation}
      \left|\mathrm{D}\psi(\boldsymbol{\xi})\right|
      =\lambda^2\frac{1}{(1-\xi_r)^2}\frac{1+\xi_z^2}{(1-\xi_z^2)^2}\,.
    \end{equation}

    In the case of a binary in the external field the domain of
    the integral is split into four subdomains with individual
    transformations:

    \noindent i) 
    \begin{multline}
      \psi_1:\, [0,100\Delta z]\;\times\;[-10\Delta z,10\Delta z]\;\to\;
          [0,100\Delta z]\;\times\;[-10\Delta z,10\Delta z]\\
          \hspace{1.8cm}(\xi_r,\xi_z) \mapsto
          \left(\xi_r,\xi_z\right)\\
          \hspace{1.8cm}\left|\mathrm{D}\psi_1(\boldsymbol{\xi})\right|=1
          \hfill
    \end{multline}
    
    \noindent ii) 
    \begin{multline}
      \psi_2:\, [0,1[\;\times\;[-10\Delta z,10\Delta z]\;\to\;
          [100\Delta z,\infty[\;\times\;[-10\Delta z,10\Delta z]\\
              \hspace{0.8cm}\left(\xi_r,\xi_z\right)
              \mapsto
              (100\Delta z + 10^7\frac{\xi_r}{1-\xi_r},\xi_z)\\
              \hspace{0.8cm}\left|\mathrm{D}\psi_2(\boldsymbol{\xi})\right|
              =10^7\frac{1}{\left(1-\xi_r\right)^2}
          \hfill
    \end{multline}

    \noindent iii) 
    \begin{multline}
      \psi_3:\, [0,1[\;\times\;[0,1[\;\to\;
          [0,\infty[\;\times\;[10\Delta z,\infty[\\
              \hspace{0.8cm}\left(\xi_r,\xi_z\right)
              \mapsto
              (10^7\frac{\xi_r}{1-\xi_r},
              10\Delta z + 10^7\frac{\xi_z}{1-\xi_z})\\
              \hspace{0.8cm}\left|\mathrm{D}\psi_3(\boldsymbol{\xi})\right|
              =10^{14}\frac{1}{\left(1-\xi_r\right)\left(1-\xi_z\right)}
          \hfill
    \end{multline}

    \noindent iv) 
    \begin{multline}
      \psi_4:\, [0,1[\;\times\;]-1,0]\;\to\;
          [0,\infty[\;\times\;[-\infty,-10\Delta z]\\
              \hspace{0.8cm}\left(\xi_r,\xi_z\right)
              \mapsto
              (10^7\frac{\xi_r}{1-\xi_r},
              -10\Delta z + 10^7\frac{\xi_z}{1+\xi_z})\\
              \hspace{0.8cm}\left|\mathrm{D}\psi_4(\boldsymbol{\xi})\right|
              =10^{14}\frac{1}{\left(1-\xi_r\right)\left(1+\xi_z\right)}
          \hfill
    \end{multline}
    In order to explore the simple transition function we here
    refrain from the use of highly efficient adaptive cubature methods. Instead,
    a two-dimensional product rule of closed Newton-Cotes methods
    is applied on an equidistant grid with incrementally decreasing
    bin size in order to force the symmetric evaluation of the
    integrand.

    \section{The QUMOND two-body system in isolation}\label{sec_qumond_isolation}
     
    \subsection{Convergence of the integral}
\begin{figure}
      \includegraphics[width=\columnwidth]{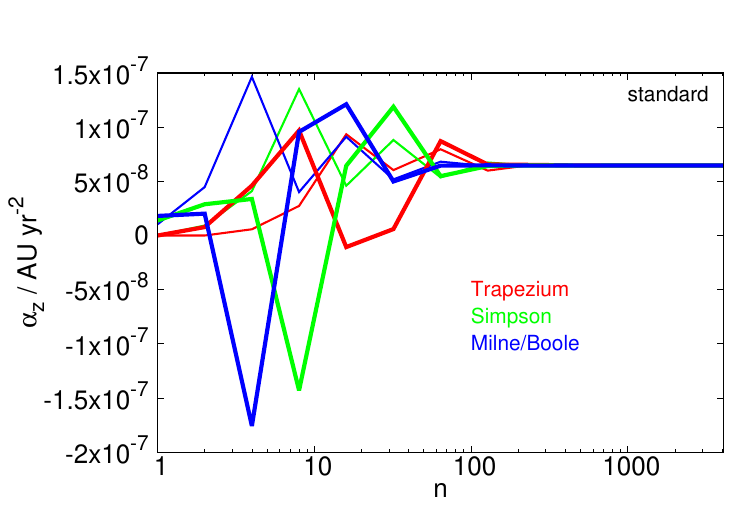}
      \caption{\label{fig_speed_conv_1}
        Speed of the convergence of the numerical integration of
        Eq.~(\ref{eq_alpha}) for
        an equal-mass isolated binary and the standard transition function.
      }
    \end{figure}

    An equal-mass binary ($m_1=m_2=1\,M_\odot$) with a separation of
    $\Delta z = 10000\,\rm AU$ is calculated with the standard transition function.
    Because both masses are equal, they have identical accelerations
    except of the sign. Figure~\ref{fig_speed_conv_1} shows the
    speed of the convergence of the absolute value of the acceleration
    as a function
    of the one-dimensional cell number for the three simplest 
    Newton-Cotes formulas.
    Particle one is located at the
    origin, $\mathbf{r}_1 = (0,0,0)$ (thin lines),
    particle 2 is positioned on the
    positive $z$-axis, $\mathbf{r}_2=(0,0,\Delta z)$
    (thick lines). The accelerations converge to the same value.
    
\begin{figure}
      \includegraphics[width=\columnwidth]{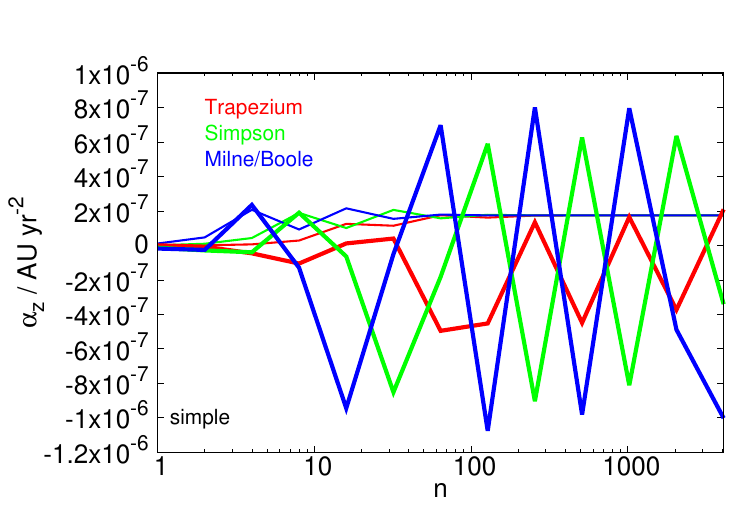}
      \caption{\label{fig_speed_conv_2}
        Speed of the (non)-convergence of the numerical integration of
        Eq.~(\ref{eq_alpha}) for
        an equal-mass isolated binary and the simple transition function
        (symmetric evaluation for particle one, non-symmetric evaluation
        for particle two).
      }
    \end{figure}

    In the case of the simple transition function the numerical integration
    converges only for particle one, which is located at the origin. The basis points
    for the function evaluation are symmetrically distributed around the particle
    and the antisymmetric part of the integrand compensates itself. The symmetric
    part is absolutely integrable and the integration converges
    (Fig.~\ref{fig_speed_conv_2}, thin lines). In the case of particle two
    the basis points of the integration routine are distributed asymmetrically 
    around $\mathbf{r}_2$ and the numerical integral does not converge
    (Fig.~\ref{fig_speed_conv_2}, thick lines).
    
\subsection{Conservation of the linear momentum}
\begin{figure}
      \includegraphics[width=\columnwidth]{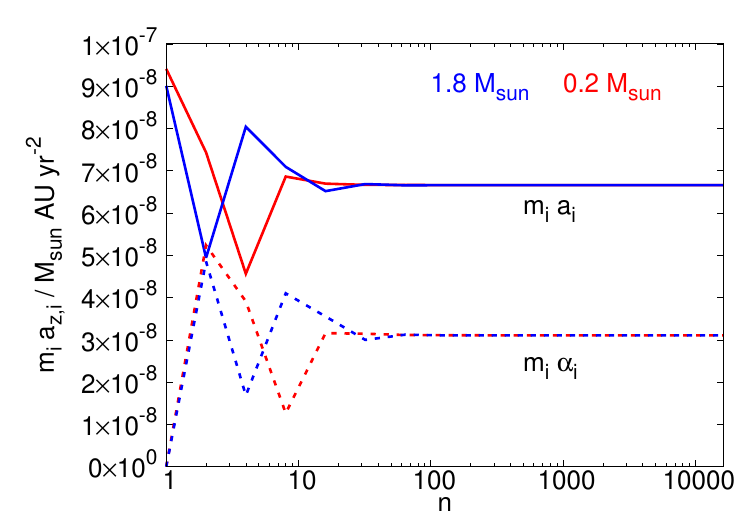}
      \caption{\label{fig_force_n}
        Forces of both components of
        a non-equal-mass isolated binary (simple transition function,
        symmetric evaluation).
      }
\end{figure}

It has been shown in \citet{milgrom2010a} that the total linear momentum
is conserved in the QUMOND field formulation and the total force vanishes.
As a test of the numerical scheme the $z$-component of
the forces acting on each particle
are calculated,
\[
F_{z,i} = m_i a_{z,i} = m_i (g_{z,i} +\alpha_{z,i})\,,
\]
and their absolute values of a non-equal-mass binary ($m_1=1.8\,M_\odot\,,m_2=0.2\,M_\odot\,,\Delta z = 10000\,\rm AU$) 
are shown in Fig.~\ref{fig_force_n} (solid lines)
as a function of the 1-d cell number (simple transition function and symmetric evaluation). 
Indeed, the absolute values of the forces are identical.
As the pure Newtonian part of the linear momentum is already conserved,
the QUMOND part of the linear momentum fulfils an individual 
conservation law,
\begin{equation}
0 = \frac{dp_{\alpha,z}}{dt} = m_1 \alpha_{z,1} + m_2 \alpha_{z,2}\,.
\end{equation}
The QUMOND part of the total force is shown in Fig.~\ref{fig_force_n} (dashed lines).

\subsection{Deep MOND limit}
In the general deep MOND limit the force between two particles with masses
$m_1$ and $m_2$ and a distance $z$ is calculated
analytically by \citep{milgrom1994a,milgrom2014a} to be
\begin{equation}\label{eq_dml_force}
  F_{2,1} = \frac{2\sqrt{Ga_0}}{3z}
  \left(
  (m_2+m_1)^{3/2}-m_1^{3/2}-m_2^{3/2}
  \right)\,,
\end{equation}
after restricting the MOND field equation to the deep MOND limit.

Considering a point mass system in deep MOND requires two limiting operations:
i) $\varepsilon$ (the radius of the symmetric domain of the Dirac sequence
tends to zero.
ii) The distance of the particles tends to infinity.

The general procedure is, first, to perform the transition of the QUMOND field equation
into the deep MOND regime and, second, to consider point masses.
This is in contrast to the procedure here: first, the accelerations of a general
$N$-body system are considered. Then, in the second step, the particles are pushed apart
from each other into the deep MOND regime.
Thus, the limits are interchanged. The minimum requirement of interchanging limits is
uniform convergence\footnote{E.g. $\lim\limits_{m\to\infty} \lim\limits_{n\to\infty}\frac{n}{x+n+m}=1$,
  interchanging the limits: $\lim\limits_{n\to\infty} \lim\limits_{m\to\infty}\frac{n}{x+n+m}=0$ }.
As the involved force laws ($1/r$ and $1/r^2$) are non-uniformly
continuous mappings, it is not clear how these procedures differ from each other.

During the transition of the QUMOND field equation into the deep MONDian
regime the Newtonian part is removed. In contrast,
in the total acceleration in Eq.~(\ref{eq_g_alpha}) the Newtonian
term does not disappear
when transiting into the deep MONDian regime. 
Thus, both procedures cannot be completely identical.
However, this term becomes negligible.

\begin{figure}
      \includegraphics[width=\columnwidth]{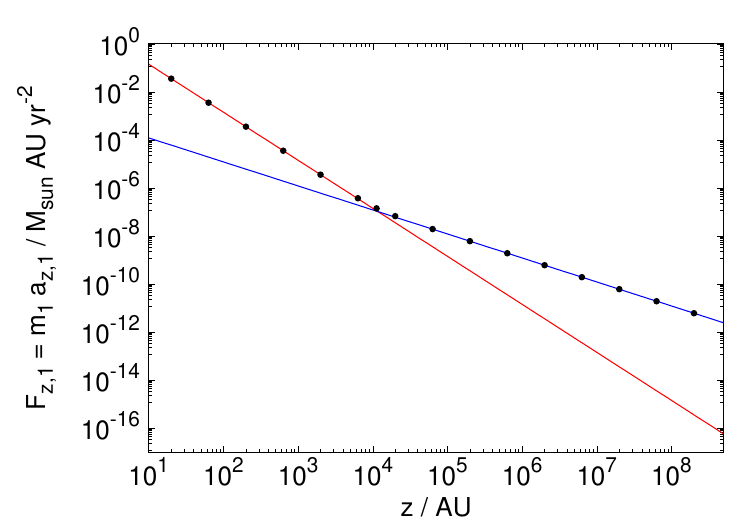}
      \caption{\label{fig_force_r}
        Total force acting on the high-mass particle as a
        function of the separation of
        the isolated binary described
        (simple transition function, symmetric evaluation).
      }
\end{figure}

Figure~\ref{fig_force_r} shows the total force acting on the high-mass 
particle ($m_1=1.8\,M_\odot\,,m_2=0.2\,M_\odot$)
as a function of the separation of the two particles from
each other. 
The numerical results are marked by black solid circles.
The analytical two-body force (Eq.~(\ref{eq_dml_force})) is
shown as the shallower solid blue line. The Newtonian force
$F_\mathrm{N} = G m_1m_2/r^2$ is indicated by the steeper red line.

\subsection{Effective one-body system}
From Eq.~(\ref{eq_dml_force}) it can be obtained that the relative
acceleration of two particles with masses $m_1$ and $m_2$ is smaller
than the acceleration of a massless test particle ($m_\mathrm{tp}=0$)
in the field of a particle of combined mass $m_\mathrm{tot} = m_1 + m_2$.

\begin{figure}
  \includegraphics[width=\columnwidth]{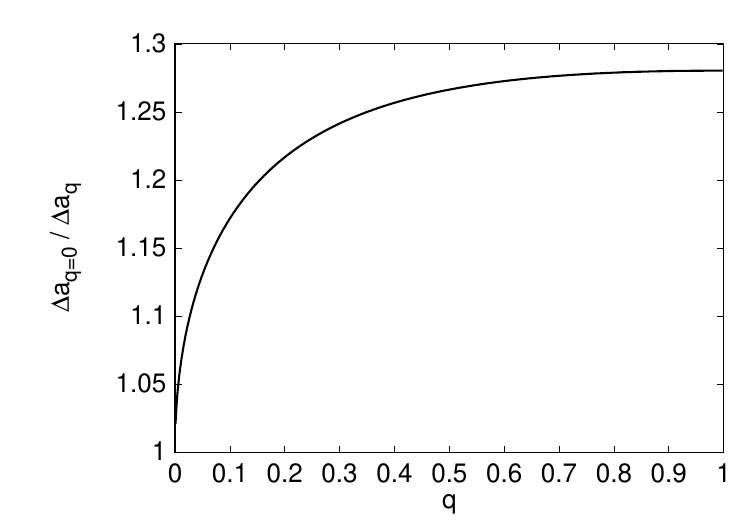}
  \caption{Boost factor of an equal-mass test particle system\label{fig_q_factor}.}
\end{figure}

In the test particle system the acceleration of the central
particle with $m_\mathrm{tot}\not=0$ is
\begin{equation}
  \begin{split}
  a_\mathrm{tot} =& \lim\limits_{m_2\to 0}\frac{F_{2,1}}{m_\mathrm{tot}} \\
  =&\lim\limits_{m_2\to 0}
  \frac{2\sqrt{Ga_0}}{3zm_\mathrm{tot}}
  \left(m_\mathrm{tot}^{3/2}-(m_\mathrm{tot}-m_2)^{3/2}-m_2^{3/2}
  \right) = 0\,.
  \end{split}
\end{equation}
The acceleration of the test particle can be calculated using L'Hôpital's rule
\begin{equation}
  \begin{split}
  a_\mathrm{tp} =& \lim\limits_{m_2\to 0}\frac{F_{2,1}}{m_2} \\
  =&\lim\limits_{m_2\to 0}
  \frac{2\sqrt{Ga_0}}{3zm_2}
  \left(m_\mathrm{tot}^{3/2}-(m_\mathrm{tot}-m_2)^{3/2}-m_2^{3/2}
  \right)\\
  =&\lim\limits_{m_2\to 0}
  \frac{\sqrt{Ga_0}}{z}
  \left((m_\mathrm{tot}-m_2)^{1/2}-m_2^{1/2}
  \right)=\frac{\sqrt{Ga_0}\sqrt{m_\mathrm{tot}}}{z}\,.
  \end{split}
\end{equation}
Inserting the mass ratio $q = m_2/m_1$ with $0\le q\le 1$, the acceleration
of the more massive particle is
\begin{equation}
  a_1 = \frac{2\sqrt{Ga_0m_\mathrm{tot}}}{3z}\frac{(1+q)^{3/2}-1-q^{3/2}}{(1+q)^{1/2}}
\end{equation}
and of the less massive particle
\begin{equation}
  a_2 = \frac{2\sqrt{Ga_0m_\mathrm{tot}}}{3z}\frac{(1+q)^{3/2}-1-q^{3/2}}{q(1+q)^{1/2}}\,.
\end{equation}
The boost factor of the internal deep MOND acceleration of
the effective test particle system
with equal masses compared to the full two-body system is given by
\begin{equation}\label{eq_q_boost}
  \frac{\Delta a_{q=0}}{\Delta a_q}
  =\frac{a_\mathrm{tot}+a_\mathrm{tp}}{a_{m_1}+a_{m_2}}
  =\frac{3q(1+q)^{1/2}}{2(1+q)\left((1+q)^{3/2}-1-q^{3/2}\right)}
\end{equation}
and depends only on the mass ratio.
The boost factor as a function of $q$ is shown in Fig.~\ref{fig_q_factor}.
The boost factor converges to 1 for $q\to 0$ and has a maximum for $q=1$
with a value of $3/(8-\sqrt{32})=1.28\ldots$.

\begin{figure}
  \includegraphics[width=\columnwidth]{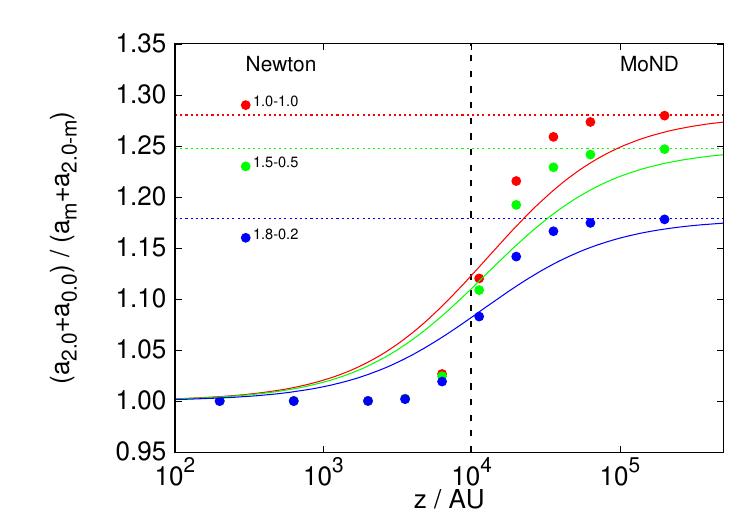}
  \caption{Boost factor of systems\label{fig_q_factor_z}
    with constant total mass $m_\mathrm{tot}=2\,M_\odot$.
    The radial evolution is shown for three different systems:
    $m_1=1.0\,M_\odot\;,\;m_2=1.0\,M_\odot$ (red dots),
    $m_1=1.5\,M_\odot\;,\;m_2=0.5\,M_\odot$ (green dots) and
    $m_1=1.8\,M_\odot\;,\;m_2=0.2\,M_\odot$ (blue dots).
    The vertical dashed line marks the MOND radius
    $r_\mathrm{M}=\sqrt{Gm_\mathrm{tot}/a_0}$. The horizontal
    dashed lines mark the respective boost factor in the deep MOND limit
    according to Eq.~(\ref{eq_q_boost}). For comparison, the solid lines
    show the boost factor obtained from  the analytical two-body force
    from \citep[their Eq.~(A2)]{zhao2010a}.}
\end{figure}

Figure~\ref{fig_q_factor_z} shows the numerically calculated 
boost factors as a function
of the separation of the two masses for different mass ratios as a function
of the separation in the case
of the standard transition function.
In the Newtonian regime the boost factor of the different configurations 
is 1, meaning that an isolated  two-body system can be replaced by
a test particle orbiting a central particle of combined mass as expected.
In the deep MOND regime the ratio of the acceleration converges very fast to a constant value following
Eq.~(\ref{eq_q_boost}).

\section{The collinear QUMOND two-body system in an external Galactic field}
\label{sec_qumond_efe}
\subsection{External field}
The aim is to construct an external field as close as possible to
the situation in \citet{banik2024a}. In their effective one-body
modelling the system is superimposed by a constant and homogenous
Newtonian gravitational field. The field is calibrated
to a circular rotation velocity of 232.8\,km/s and a Galactocentric
distance of $R_0=8200$\,pc \citep{mcmillan2017a}. These values
determines the acceleration ratio $a/a_0 = 1.785$. Due to the problem
of the replacement of a large number of particles by a smooth mass
distribution as outlined in the final paragraph of
Sect.~\ref{sec_qumond_nbody}, the external field is realised by
a central massive particle. \citet{banik2024a} used the simple transition
function. In this case the ratio of the external Newtonian acceleration
$g_e$ and the threshold acceleration $a_0$ is $g_e/a_0=1.144$ and requires
a central massive particle of
$m_\mathrm{G} = R_0^2 g_e/G=6.62\times 10^{10}\,M_\odot$.

\subsection{Distribution of the phantom dark matter}
    \label{sec_iso_pdm}
\begin{figure}
      \includegraphics[width=\columnwidth]{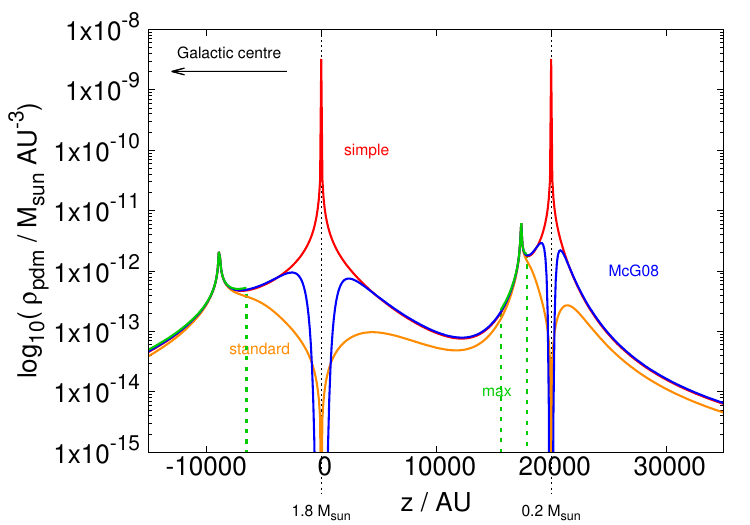}
      \includegraphics[width=\columnwidth]{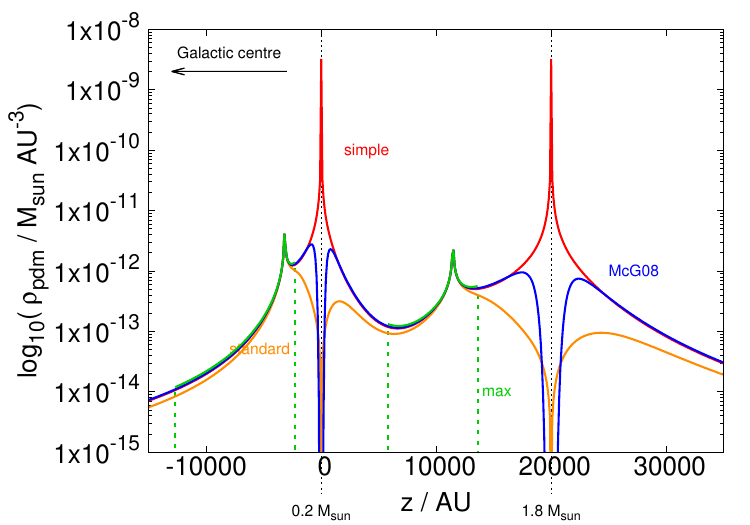}
      \caption{\label{fig_pdm_efe}
        Phantom dark matter distribution of an two-body system in the Galactic
        EFE.}
    \end{figure}

    The distribution of the phantom dark matter density along 
    the connecting line ($z$-axis) of two particles with masses
    $m_1=1.8\,M_\odot$ and $m_2=0.2\,M_\odot$ and a separation of
    $\Delta z = 20\,\rm kAU$ can be seen in Fig.~\ref{fig_pdm_efe}
    for four different transition functions (Table~\ref{tab_nu_fcts})
    and both arrangements.
    The Galactic point mass is put at $z=-8.2\,\rm kpc$.
    Now, the location of the Galactic zero-g point between both
    binary components depends on the order and there appears a second
    zero-g point between the Galactic centre and the binary.
    A massless test particle does not produce a phantom dark matter peak between itself
    and any other real particle.
    
\subsection{MONDian Path of a Newtonian binary}
\begin{figure}
      \includegraphics[width=\columnwidth]{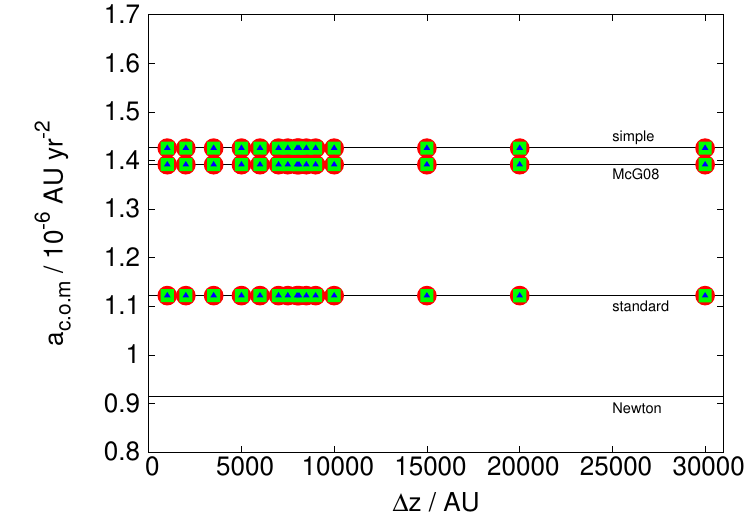}
      \caption{\label{fig_com_efe}
        Centre of mass accelerations as a function of the internal separation.
        Red circles: particle order is $m_\mathrm{gal} - 2\,M_\odot - 0\,M_\odot$.
        Green squares: particle order is $m_\mathrm{gal} - 1\,M_\odot - 1\,M_\odot$.
        Black triangles: particle order is $m_\mathrm{gal} - 0\,M_\odot - 2\,M_\odot$.
        Equal-mass binary is Newtonian for $\Delta z = \sqrt{2\,M_\odot\,G/a_0}\le 10038\,\rm AU$.}
    \end{figure}

The Newtonian external acceleration at the location of the binary is
$g_e = 9.138\times 10^{-7}\,\rm AU/yr^2$. This
corresponds to a kinematical 
acceleration of $1.426\times 10^{-6}\,\rm AU/yr^2$ (simple transition function).
Keeping the Newtonian acceleration constant and varying the transition function
the corresponding  kinematical 
accelerations are $1.122\times 10^{-6}\,\rm AU/yr^2$ (standard transition function) and 
$1.391\times 10^{-6}\,\rm AU/yr^2$ (McG08 transition function).
The acceleration of the centre of mass of the binary is calculated with
\begin{equation}
  a_{z,com} = \frac{1}{m_1+m_2}
  \left(
  m_1(g_{z,1}+\alpha_{z,1})+m_2(g_{z,2}+\alpha_{z,2})
  \right)\,.
\end{equation}
Figure~\ref{fig_com_efe} shows the numerically obtained acceleration
values of the centre of mass of the equal-mass binary ($m_1=m_2=1\,M_\odot$)
and of the effective single-body problems ($m_1=2\,M_\odot$, $m_2 = 0$)
compared to the external
Newtonian acceleration scaled by the boost factors of the different transition
functions. It can be seen that the centre of masses of the binaries move on
a MONDian binary whether or not the binary is internally MONDian or
Newtonian.

\subsection{Effective one-body system vs. two-body system}
\begin{figure}
  \includegraphics[width=\columnwidth]{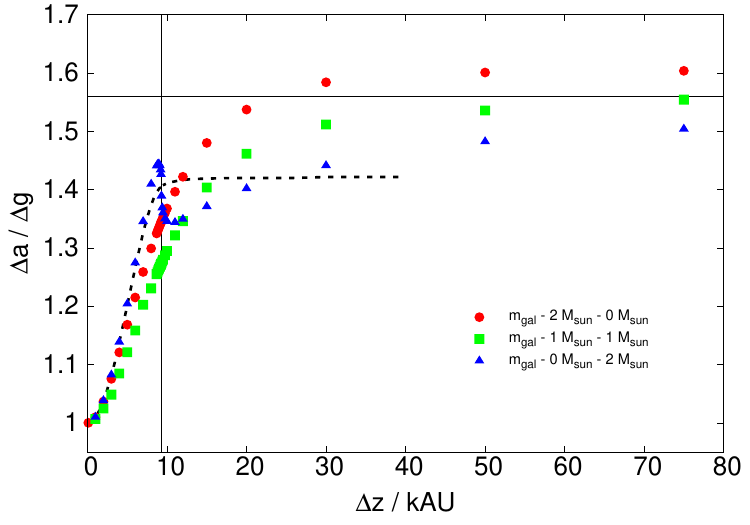}
  \includegraphics[width=\columnwidth]{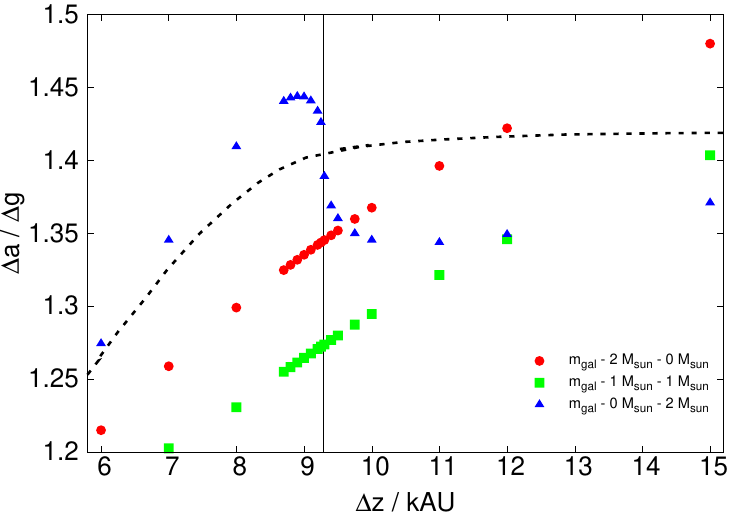}
      \caption{\label{fig_da-dg_efe}
        MOND boost factor of the internal relative acceleration in the case of the
        simple transition function.
        The horizontal solid line in the upper diagram shows the maximum boost
        $\eta_\mathrm{max} = 1.56$. The vertical solid line marks the zero-g
        distance of a massless test
        particle. The short dashed line shows the
        azimuthally averaged boost factor from
        \citet{banik2024a}. The bottom diagram shows a zoom-in
        of the upper diagram in the transition region.}
\end{figure}

Figure~\ref{fig_da-dg_efe} shows the ratio of the internal relative kinematical acceleration
to the internal relative Newtonian acceleration,
\begin{equation}
  \frac{\Delta a}{\Delta g} = \frac{|a_{z,2}-a_{z,1}|}{|g_{z,2}-g_{z,1}|}\,,
\end{equation}
as a function of the internal separation for the equal-mass two-body system and
for the effective one-body system. In the Newtonian
regime $\lesssim 10\,\rm kAU$
the boost factor of the equal-mass binary
increases slower with increasing separation than the boost factor of
both effective one-body systems in the case of the simple transition function with symmetric
evaluation.
The one-body system with the test particle between the Galactic centre and the massive binary
component shows the fastest increasement of the boost factor until the massless
test particle reaches the Galactic zero g position, where both accelerations induced
by the Galactic centre particle and the massive binary particle balance each other
($\Delta z = \sqrt{m_2/m_\mathrm{gal}}R_0 = 9293.6\,\rm AU$). Here, the boost factor
drops abruptly and shows the slowest increasement of the boost factor in the EFE-regime.
For large separations the boost factor of the equal-mass binary converges against
the theoretically expected maximum boost in the EFE-effect
($\eta_\mathrm{max} = \nu_\mathrm{simp}(g_\mathrm{ext}/a_0)=1.56$\,).
For comparison to the work by \citet{banik2024a} the azimuthally averaged boost factor
(their Figure 7) is superimposed scaled to a MOND radius of
$r_\mathrm{M} = \sqrt{G 2\,M_\odot /a0}=10038\,\rm AU$.

The unexpected behaviour of the massless test particle is due to the additional
influence of the phantom dark matter spike produced by the Galactic zero-g
point between the Galactic centre particle and the massive binary component.
When the test particle recedes from the massive component approaching the
phantom dark matter spike it feels an increasing additional attraction
towards the Galactic centre and the difference in the kinematical
acceleration increases. Once the test particle has passed the zero-g point
the additional attraction has switched the direction and leads to a fast
decrease of the difference of the accelerations.

\begin{figure}
  \includegraphics[width=\columnwidth]{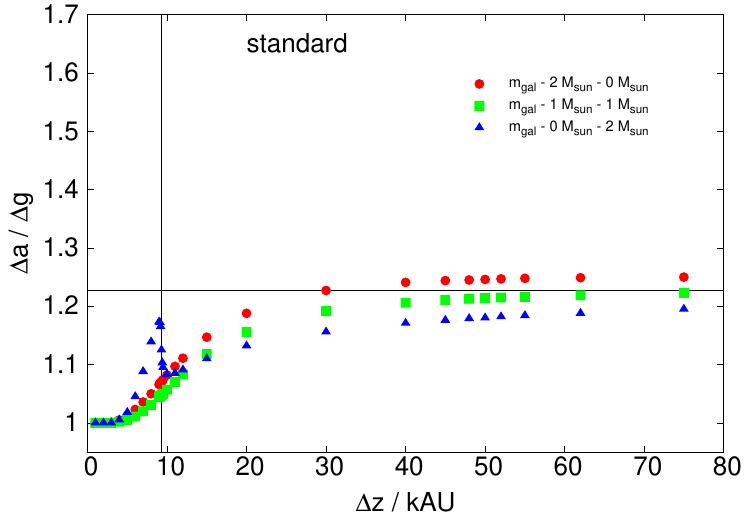}
  \includegraphics[width=\columnwidth]{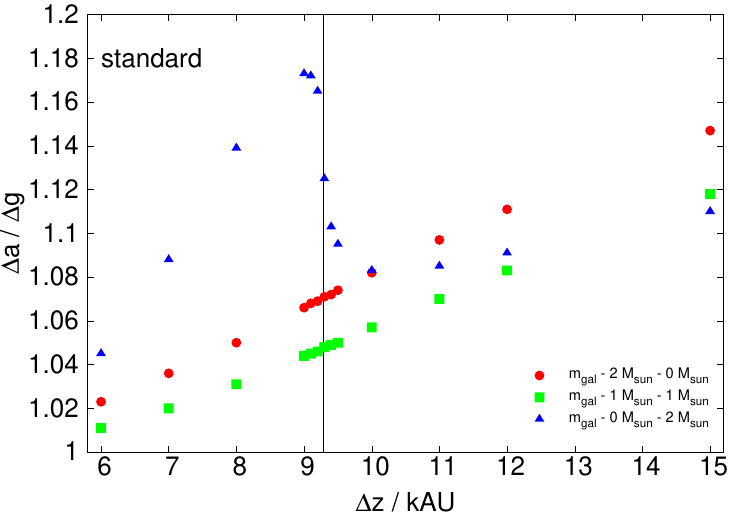}
      \caption{\label{fig_da-dg_efe_standard}
        MOND boost factor of the internal relative acceleration in the case of the
        standard transition function.
        The horizontal line in the upper diagram shows the maximum boost
        $\eta_\mathrm{max} = 1.23$.
        The vertical line marks the zero g distance of a massless test
        particle. The bottom diagram shows a zoom-in
        of the upper diagram in the transition region.}
\end{figure}

\begin{figure}
  \includegraphics[width=\columnwidth]{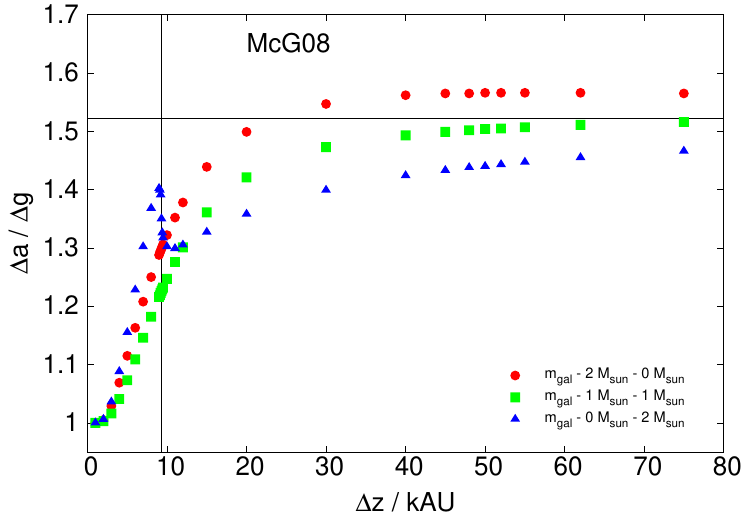}
  \includegraphics[width=\columnwidth]{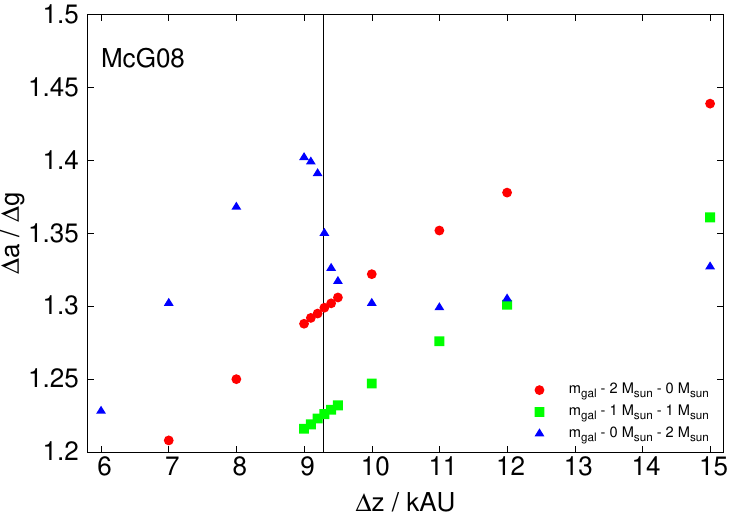}
      \caption{\label{fig_da-dg_efe_smc16}
        MOND boost factor of the internal relative acceleration in the case of the
        McG08 transition function.
        The horizontal line in the upper diagram shows the maximum boost
        $\eta_\mathrm{max} = 1.52$.
        The vertical line marks the zero g distance of a massless test
        particle. The bottom diagram shows a zoom-in
        of the upper diagram in the transition region.}
\end{figure}

The differences in the boost factors in the case of the standard transition
function is shown in Fig.~\ref{fig_da-dg_efe_standard} and in the case
of the McG08 transition function in Fig.~\ref{fig_da-dg_efe_smc16}.
All three models have in common that within the MOND radius the effective
one-body systems have higher boost factors than the equal-mass system.

\begin{figure}
  \includegraphics[width=\columnwidth]{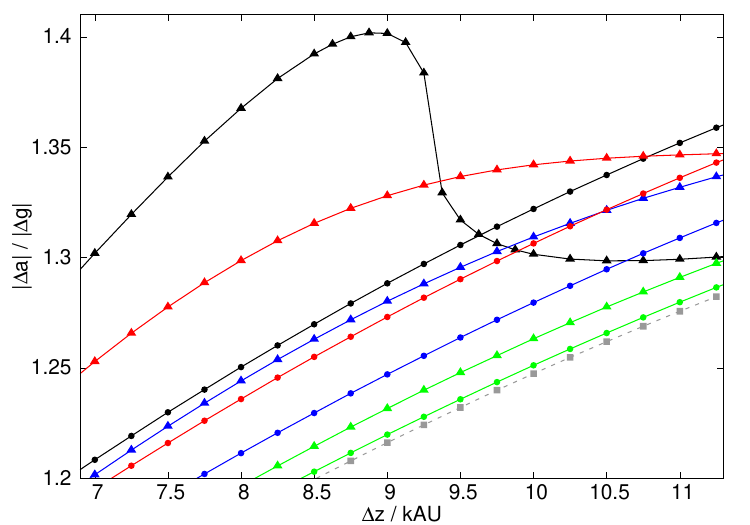}
  \caption{\label{fig_efe_q}Mass dependence of the boost factor.
    Shown is the boost factor of the collinear configuration of a binary with a
    total mass of $2\,M_\odot$ with different mass ratios as a function
    of the internal separation:
    $2.0\,M_\odot$-$0.0\,M_\odot$ (black),$1.5\,M_\odot$-$0.5\,M_\odot$ (green),
    $1.9\,M_\odot$-$0.1\,M_\odot$ (blue), $1.99\,M_\odot$-$0.01\,M_\odot$ (red).
    The equal-mass binary is plotted with grey squares. The configurations
    with the less-mass component between the higher-mass component and the
    Galactic centre particle are indicated by triangles, those
    where the less-mass component is on the side of the Galactic anti-centre
    are indicated by circles.
  }
\end{figure}

The boost factor in the transition
regime is shown in Fig.~\ref{fig_efe_q} for different mass ratios.
The equal-mass binary has got the lowest boost factor. And for all
cases configurations where the lower-mass component is located between
the Galactic centre and the higher-mass component have higher
internal relative accelerations than in case of the opposite mass
ordering.

\section{Embedded binary in three dimensions}\label{sec_3d}
In this section the embedded equal-mass binary and the corresponding effective
one-body system are considered in a three dimensional configuration for the cases
of a separation of $\Delta z = 9000\,$AU, where the discrepancy in the relative
acceleration is largest, and a separation of $\Delta z = 75000\,$AU, where the
binary is in the deep EFE. The McG08-transition function is used for these calculations. Here, the integral in Eq.~(\ref{eq_alpha}) is evaluated in three dimensions. The partitioning of the domain has been adjusted.

The mass of the first particle is kept at the origin ($m_1=2\,M_\odot$ in case of the
effective one-body system). The second mass is rotated  with a  constant
distance to the first particle
in steps of $\Delta\vartheta = 1^\circ$ in the $x-z$-plane
($x_2 =\Delta z \sin(\vartheta)$, $y_2=0$,
$z_2 =\Delta z \cos(\vartheta)$). In the case of the effective binary $\vartheta=0$
corresponds to the collinear ordering $m_\mathrm{gal}-2\,M_\odot-0\,M_\odot$
and $\vartheta=180^\circ$ corresponds
to the collinear ordering $m_\mathrm{gal}-0\,M_\odot-2\,M_\odot$ examined 
in the previous section.

\begin{figure}
  \includegraphics[width=\columnwidth]{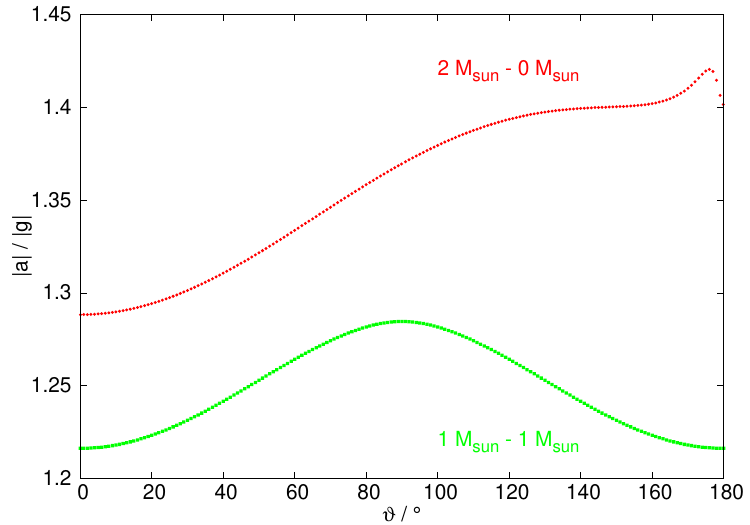}
  \caption{\label{fig_tb_angle}3d-boost factor for  $\Delta z =9000\,\rm AU$. Shown are the ratios of the
    absolute values of the total kinematical and the Newtonian acceleration as a function
    of the inclination angle for the equal-mass binary and the effective one-body system.}
\end{figure}

Figure~\ref{fig_tb_angle} shows the boost factor $|\mathbf{a_2}-\mathbf{a}_1|/|\mathbf{g_2}-\mathbf{g}_1|$ as a function of the inclination angle $\vartheta$.
In the case of the equal-mass binary the boost factor is maximal for the perpendicular
configuration and has a minimum for the collinear setting. In the case of the effective
binary the boost factor has a minimum for the configuration where the test particle
is at the outer most position and increases continuously with increasing inclination.
Interestingly, the maximum is not reach at the position where the test particle is
between the mass particle and the Galactic centre but close to it.
This might be due to the axis-symmetry of the distribution of the
phantom dark matter. If the test particle is positioned on the $z$-axis
a higher fraction of the acceleration cancels each other than if the
test particle is slightly displaced from the $z$-axis.

\begin{figure}
  \includegraphics[width=\columnwidth]{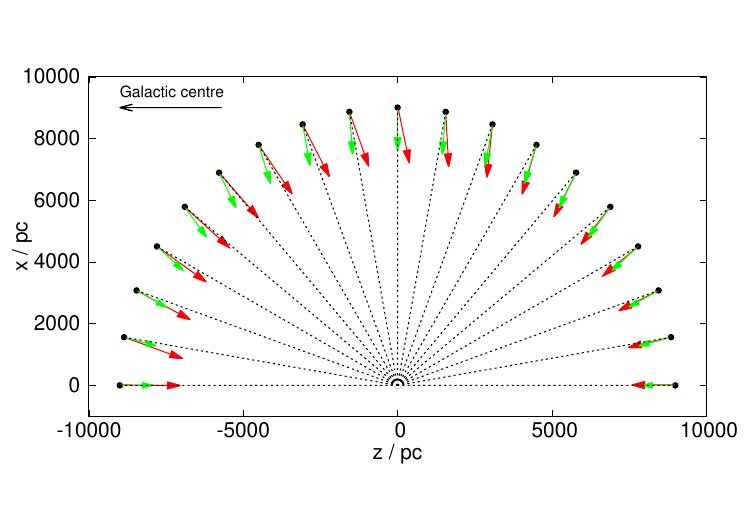}
  \caption{\label{fig_fig_tb_x_z}Directions of QUMOND contributions for
    $\Delta z$=9000~AU. The green arrows indicate the direction of the
    QUMOND contribution to the acceleration in the case of the equal-mass
    binary. The red arrows indicate the direction of the
    QUMOND contribution to the acceleration in the case of the effective
    one-body system. The length of the arrows scale linearly with the
    absolute value of the QUMOND contribution.
  }
\end{figure}

Figure~\ref{fig_fig_tb_x_z} shows different directions of the QUMOND
contribution to the kinematical acceleration.

\begin{figure}
  \includegraphics[width=\columnwidth]{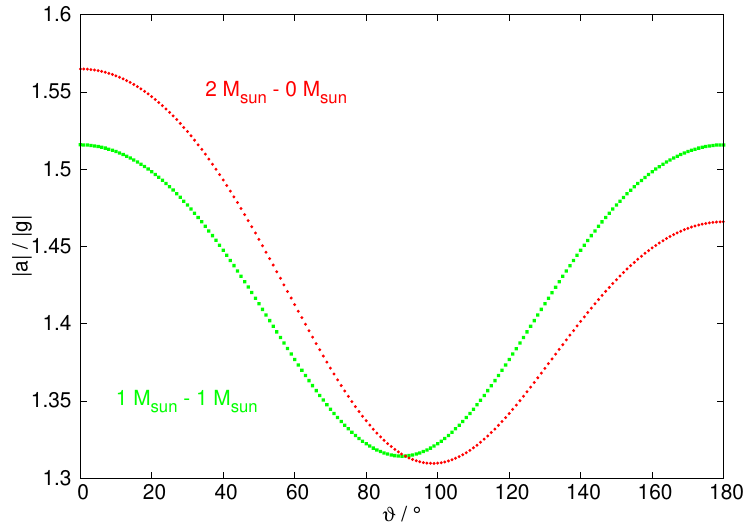}
  \caption{\label{fig_tb_angle_75000}3d-boost factor for
    $\Delta z = 75000\,\rm AU$.
    Similar to Fig.~\ref{fig_tb_angle}.}
\end{figure}

\begin{figure}
  \includegraphics[width=\columnwidth]{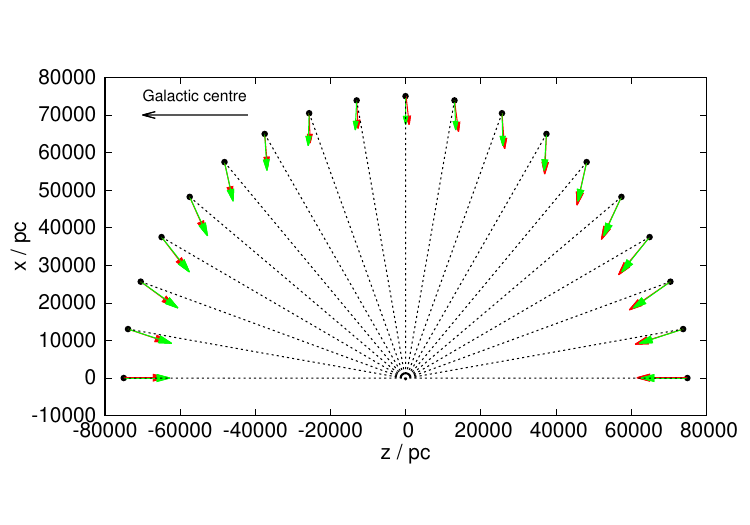}
  \caption{\label{fig_fig_tb_x_z_75000}Directions of QUMOND contributions
    for $\Delta z$=75000~AU. Similar to Fig.~\ref{fig_fig_tb_x_z}.}
\end{figure}

The boost factor for a separation of 75000~AU is shown
in Fig.~\ref{fig_tb_angle_75000}. In the deep EFE the relative acceleration
is smallest for the perpendicular orientation and largest for
the collinear configuration. The directions of the QUMOND contributions
are displayed in Fig.~\ref{fig_fig_tb_x_z_75000}.

\section{Wide binaries in Milgrom law dynamics (MLD)}\label{sec_mld}
As a different MOND formulation we here follow MLD in order to explore
the boost factors of the internal binary dynamics.
In MLD the relation between the radial
kinematical acceleration, $a_r$, and the radial Newtonian
gravitational acceleration, $g_r$, proposed by \citet{milgrom1983a}
for disk galaxies has been postulated to be valid in vectorial
form \citep{pflamm-altenburg2025a}. Although it allows an easy and fast
calculation of MONDian accelerations in $N$-body systems,
this MOND formulation is incomplete.
The disadvantage of this formulation is, that an internally Newtonian
binary follows the external Newtonian path rather than the MONDian orbit.
In order to suppress the Newtonisation of compact subsystems the Newtonian
gravitational force has been softened with a smoothing parameter of 0.001~pc,
which allows to cover the full internal MONDian regime of wide binaries
and open star clusters. It has been demonstrated in
\citet{pflamm-altenburg2025a} that contrary to previous statements \citep{felten1984a}
an isolated MONDian two-body system does not show a general self-acceleration in the
meaning of getting continuously faster. The constant
and uniform motion of the Newtonian
centre of mass is replaced by a constant and uniform motion of the MONDian
centre of mass.

\begin{figure}
      \includegraphics[width=\columnwidth]{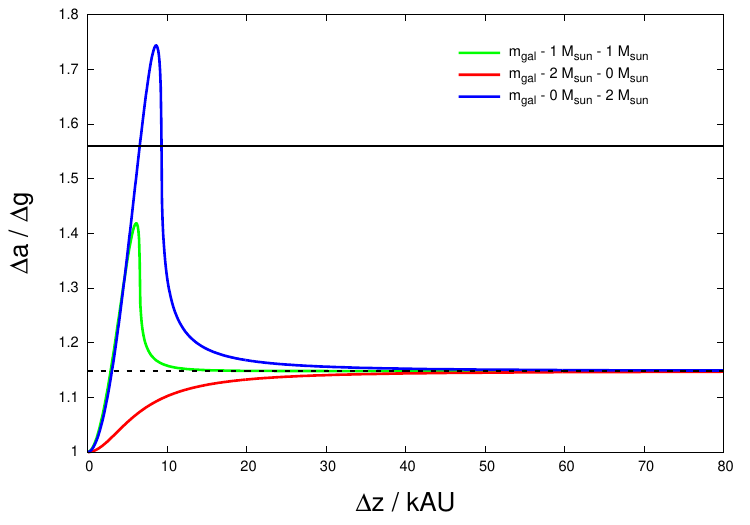}
      \caption{\label{fig_da-dg_mld_simp}
        Boost factor in MLD with simple transition function. The solid horizontal line
        shows the wrong boost factor (Eq.~(\ref{eq_mld_wrong})).
      The short dashed horizontal line
        shows the correct boost factor (Eq.~(\ref{eq_mld_correct})).}
\end{figure}

In MLD the kinematical acceleration along the $z$-axis of particle $i$ is
\begin{equation}
  a_i = \nu(|g_i+g_e|/a_0)\,(g_i+g_e)\,,
\end{equation}
where now the external acceleration is assumed to be homogeneous and
 $g_i$ is the
Newtonian acceleration exerted by the other binary component.
In a simple estimate one might argue that for large internal separation the $\nu$-factor
is dominated by the external field ($\nu(g_e/a_0)$) and the boost factor
is
\begin{equation}\label{eq_mld_wrong}
  \frac{\Delta a}{\Delta g} = \frac{|a_2-a_1|}{|g_2-g_1|}
  =\nu(g_e/a_0)\frac{|g_2-g_1|}{|g_2-g_1|} = \nu(g_e/a_0)>1\,.
\end{equation}
Taking the same external Newtonian acceleration as in the previous section
this boost factor is $\nu_\mathrm{simple}(9.138\times 10^{-7}\,\rm AU\,yr^{-2}/a_0) = 1.56$
in the case of the simple transition function. Figure~\ref{fig_da-dg_mld_simp} shows
the numerical evaluation of the boost factor in MLD as a function of the internal
separation and is much below the estimated value of 1.56.

Expanding the transition function into a Taylor series an analysis of the
limit leads to the correct boost factor of
\begin{equation}\label{eq_mld_correct}
  \lim\limits_{\Delta z \to \infty} \frac{|a_2-a_1|}{|g_2-g_1|}
  = \nu(g_e/a_0) + \nu^\prime(g_e/a_0)\,g_e/a_0\,.
\end{equation}
For the simple transition function the true boost factor is 1.15,
as can be seen in Fig.~\ref{fig_da-dg_mld_simp}. In contrast to the QUMOND
formulation the boost factor of the equal-mass binary lies between both
effective one body systems.

\begin{figure}
      \includegraphics[width=\columnwidth]{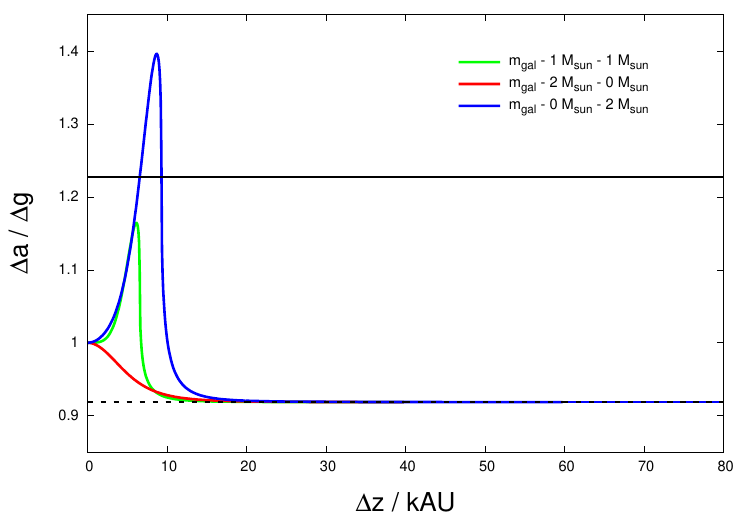}
      \caption{\label{fig_da-dg_mld_std}
       Boost factor in MLD with standard transition function.}
\end{figure}

Each transition function has got values larger than one
and a negative derivative. 
Thus, boost factors smaller than one are possible in MLD.
For the standard transition function
the wrong boost factor is $1.23$ and the correct one
is $0.92$ (Fig.~\ref{fig_da-dg_mld_std}). MLD allows a reduction
of the internal dynamics, depending on the transition function
and the strength of the external field.

\section{Proposal of an observational test}\label{sec_proposal}
\begin{figure}
      \includegraphics[width=\columnwidth]{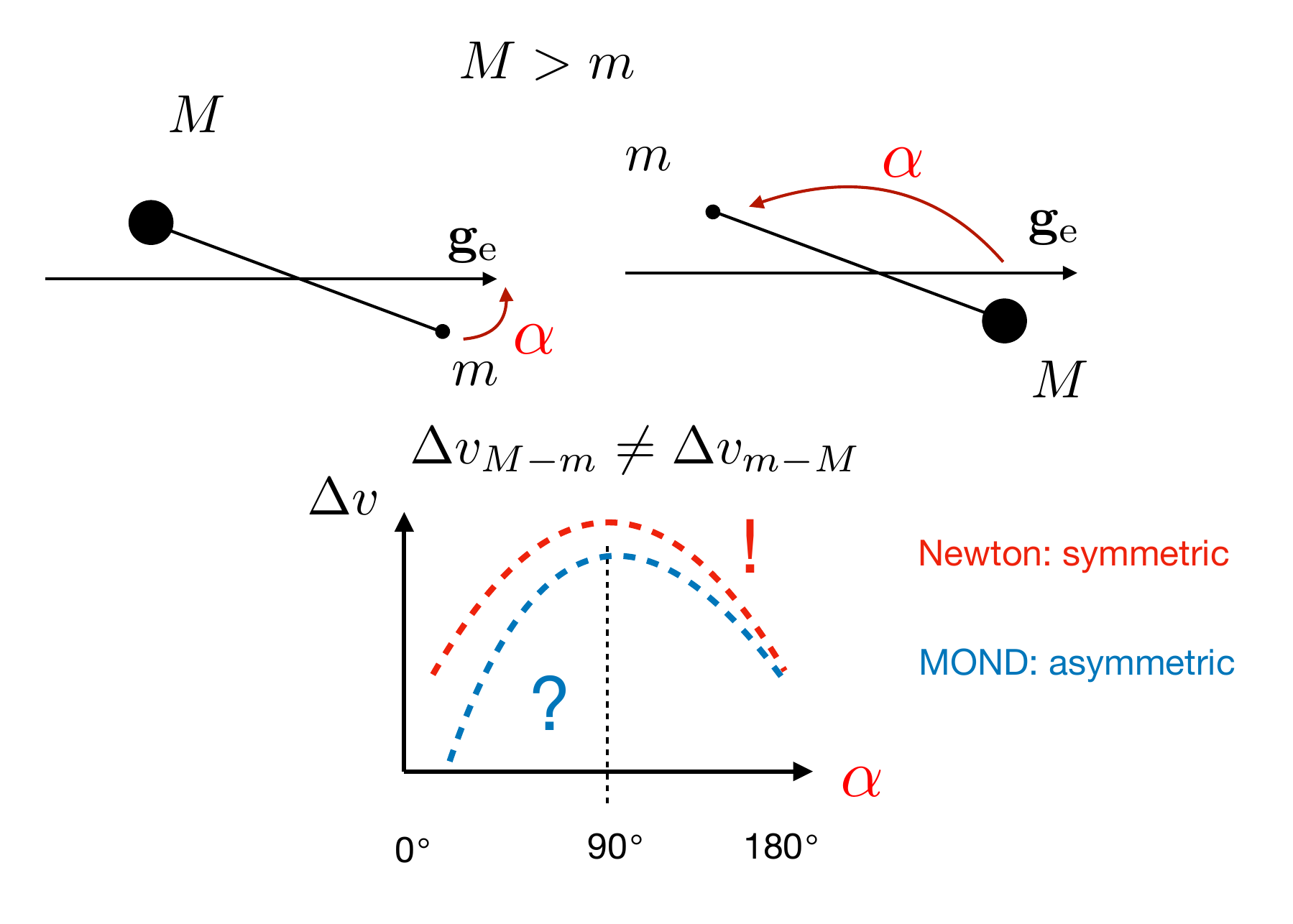}
      \caption{\label{fig_delta_v}
       Expected asymmetry of relative velocities.}
\end{figure}

All tests on wide binaries compare the set of observed data pairs
of the projected relative velocity with the projected separation
of both constituents with those obtained from Newtonian
and non-Newtonian modelling of wide binary populations.
The detailed modelling of an ensemble of binaries require
in addition to the gravitational theory assumptions on
the distribution of structural parameters such as for example
the mass ratio, the eccentricity and the semi major axis.
However, a fundamental difference between Newtonian and
MONDian binaries is that the relative acceleration  between
both constituents depends in MOND on the inclination angle of both stars
in the case they have non-equal masses. The relative velocity
between both stars should have an asymmetrical dependence
on the position angle of the binary on sky with respect to the
direction of the external field.
This means, it depends on whether or not the more massive component
is closer to the Galactic centre (Fig.~\ref{fig_delta_v}).

But given the present line-of-sight errors the three-dimensional
resolution of the binary configurations is too poor and such
an observational test is currently not feasible. An observational
detection of the effect is even more complicated, as demonstrated
in the previous section, because the difference in the accelerations
are only obvious for large mass ratios which are untypical for stellar binaries.

However, the asymmetry in the acceleration and velocity field should exist
in the Solar system as well. Thus, Oort-cloud and trans-Neptunian objects
are also candidates to search for non-Newtonian asymmetries as their have
very large mass ratios with respect to the sun.
Despite the current observational impossibility it is worth to
work out detailed predictions in QUMOND dynamics and other MOND formulations.

\section{Summary and Conclusions}
Here, we constructed a method to obtain the acceleration of particles
in an $N$-body system in QUMOND, which is consistent with the field formulation
of QUMOND. The acceleration of a point mass is expressed by a Green's
integral, obtained by treating point masses as limits of Dirac sequences
just as the Newtonian acceleration of a point mass is
obtained from the Poisson equation.

All QUMOND-transition functions ensure integrability at infinity
and at zero-g points. At the particle singularity the simple transition
function is not absolutely integrable, whereas the McGaugh and the standard
transition function are.

This method is first applied to an isolated binary and compared to
analytical solutions in the deep MOND limit. In the second step
a wide binary embedded in an external Galactic field similar to
the solar vicinity is explored. If a binary is approximated by an
effective one-body system where the central particle contains
the total system mass and is orbited by a massless test particle
the relative acceleration is larger than in the full two-body system
in the transition regime.

This has an important consequence for the results in \citet{banik2024a}.
As the vast majority of the binaries lie in the transition regime the
obtained accelerations in their modelling are larger than in the correct
handling of binaries in QUMOND. Thus, in order to test binary dynamics
in QUMOND against observational data requires the development of 
proper and consistent QUMOND-$N$-body orbit integrators.
However, to speed up the
calculations to a tolerable duration this requires
to replace the three-dimensional tensor-product integration methods
by an efficient non-tensor-type integration algorithm with an
suitable adaptive subpartitioning scheme in order to resolve the particle
singularities and the phantom dark matter spikes at the Galactic zero-g
points.

An additional zero-g point between both binary components appears.
As these zero-g points show a local peak of phantom dark matter,
it is worth to explore weather or not baryonic matter can be trapped  
and if this is in principle detectable in the solar system.

In this first step the collinearly embedded binary is considered
because the numerical integration simplifies by reducing the integration
from three to two dimensions. It is now straight forward to
extend the numerical integration to three dimensions and to reinvestigate
in more detail the orbit of Saturn
in the context of the Cassini ranging data
\citep{blanchet2011a,hees2014a,hees2016a},
the effective planet nine hypothesis \citep{pauco2016a,pauco2017a,
brown2023a} and the wide-binary orbit modelling.

Simultaneously to the investigation of QUMOND $N$-body systems
it is desired to construct expressions for particle accelerations
in $N$-body systems which are fully consistent
with the field formulation of AQUAL. However, first considerations
show that this involves the numerical solution of 
three dimensional non-linear Fredholm integral
equations of second kind.

\bibliography{ms}{}
\bibliographystyle{aa}

\end{document}